\documentclass[onecolumn,11pt]{article}
\usepackage[top=1in, bottom=1in, left=1in, right=1in]{geometry}
\usepackage{amsmath,amsfonts,amscd,amssymb,amsthm}
\usepackage{graphicx}
\usepackage{epstopdf}
\usepackage{mathtools}
\usepackage{setspace}
\usepackage{palatino} 
\usepackage[numbers,sort&compress]{natbib}
\usepackage{bm}

\usepackage[bottom,flushmargin,hang,multiple]{footmisc}
\usepackage{lipsum}
\newcommand\blfootnote[1]{%
  \begingroup
  \renewcommand\thefootnote{}\footnote{#1}%
  \addtocounter{footnote}{-1}%
  \endgroup
}

\title{{\fontsize{16}{16}\selectfont \textbf{Applying Machine Learning to Study Fluid Mechanics}}\vspace{-.15in}}

\author{\normalsize{Steven L. Brunton$^{1*}$}\\
\footnotesize{$^1$ Department of Mechanical Engineering, University of Washington, Seattle, WA 98195, United States\vspace{-.2in}}
}
\date{}
\graphicspath{/figures}
\begin{document}
\maketitle

\blfootnote{$^*$ Corresponding author (sbrunton@uw.edu).}
\vspace{-.2in}
\begin{abstract}
This paper provides a short overview of how to use machine learning to {build data-driven models} in fluid mechanics.  
The process of machine learning is broken down into five stages: (1) formulating a problem to model, (2) collecting and curating training data to inform the model, (3) choosing an architecture with which to represent the model, (4) designing a loss function to assess the performance of the model, and (5) selecting and implementing an optimization algorithm to train the model.  
At each stage, we discuss how prior physical knowledge may be embedding into the process, with specific examples from the field of fluid mechanics.  

\noindent\emph{Keywords--}
Machine learning, fluid mechanics, physics-informed machine learning, neural networks, deep learning\end{abstract}

\section{Introduction}

The field of fluid mechanics is rich with data and rife with problems, which is to say that it is a perfect playground for machine learning. 
Machine learning is the art of building models from data using optimization and regression algorithms.  
Many of the challenges in fluid mechanics may be posed as optimization problems, such designing a wing to maximize lift while minimizing drag at cruise velocities, estimating a flow field from limited measurements, controlling turbulence for mixing enhancement in a chemical plant or drag reduction behind a vehicle, among myriad others. 
These optimization tasks fit well with machine learning algorithms, which are designed to handle nonlinear and high-dimensional problems. 
In fact, machine learning and fluid mechanics both tend to rely on the same assumption that there are patterns that can be exploited, even in high-dimensional systems~\cite{Taira2017aiaa}. 
Often, the machine learning algorithm will model some aspect of the fluid, such as the lift profile given a particular airfoil geometry, providing a \emph{surrogate} that may be optimized over. 
Machine learning may also be used to directly solve the fluid optimization task, such as designing a machine learning model to manipulate the behavior of the fluid for some engineering objective with active control~\cite{rabault2019artificial,ren2020active,zhou2020artificial}.  

In either case, it is important to realize that machine learning is \emph{not} an automatic or turn-key procedure for extracting models from data.  
Instead, it requires expert human guidance at every stage of the process, from deciding on the problem, to collecting and curating data that might inform the model, to selecting the machine learning architecture best capable of representing or modeling the data, to designing custom loss functions to quantify performance and guide the optimization, to implementing specific optimization algorithms to train the machine learning model to minimize the loss function over the data.  
A better name for machine learning might be ``expert humans teaching machines how to learn a model to fit some data," although this is not as catchy.  
Particularly skilled (or lucky!) experts may design a learner or a learning framework that is capable of learning a variety of tasks, generalizing beyond the training data, and mimicking other aspects of intelligence.  
However, such artificial intelligence is rare, even more so than human intelligence.  
The majority of machine learning models are just that, models, which should fit directly into the decades old practice of model-based design, optimization, and control~\cite{Brunton2020arfm}.  

With its unprecedented success on many challenging problems in computer vision and natural language processing, machine learning is rapidly entering the physical sciences, and fluid mechanics is no exception. 
The simultaneous promise, and over-promise, of machine learning is causing many researchers to have a healthy mixture of optimism and skepticism.  
In both cases, there is a strong desire to understand the uses and limitations of machine learning, as well as best practices for how to incorporate it into existing research and development workflows. 
It is also important to realize that while it is now relatively simple to train a machine learning model for a well-defined task, it is still quite difficult to create a new model that outperforms traditional numerical algorithms and physics-based models.  
{Incorporating partially known physics into the machine learning pipeline well tend to improve model generalization and improve interpretability and explainability, which are key elements of modern machine learning~\cite{du2019techniques,molnar2020interpretable}.  }

\begin{figure}
    \centering
    \includegraphics[width=.7\textwidth]{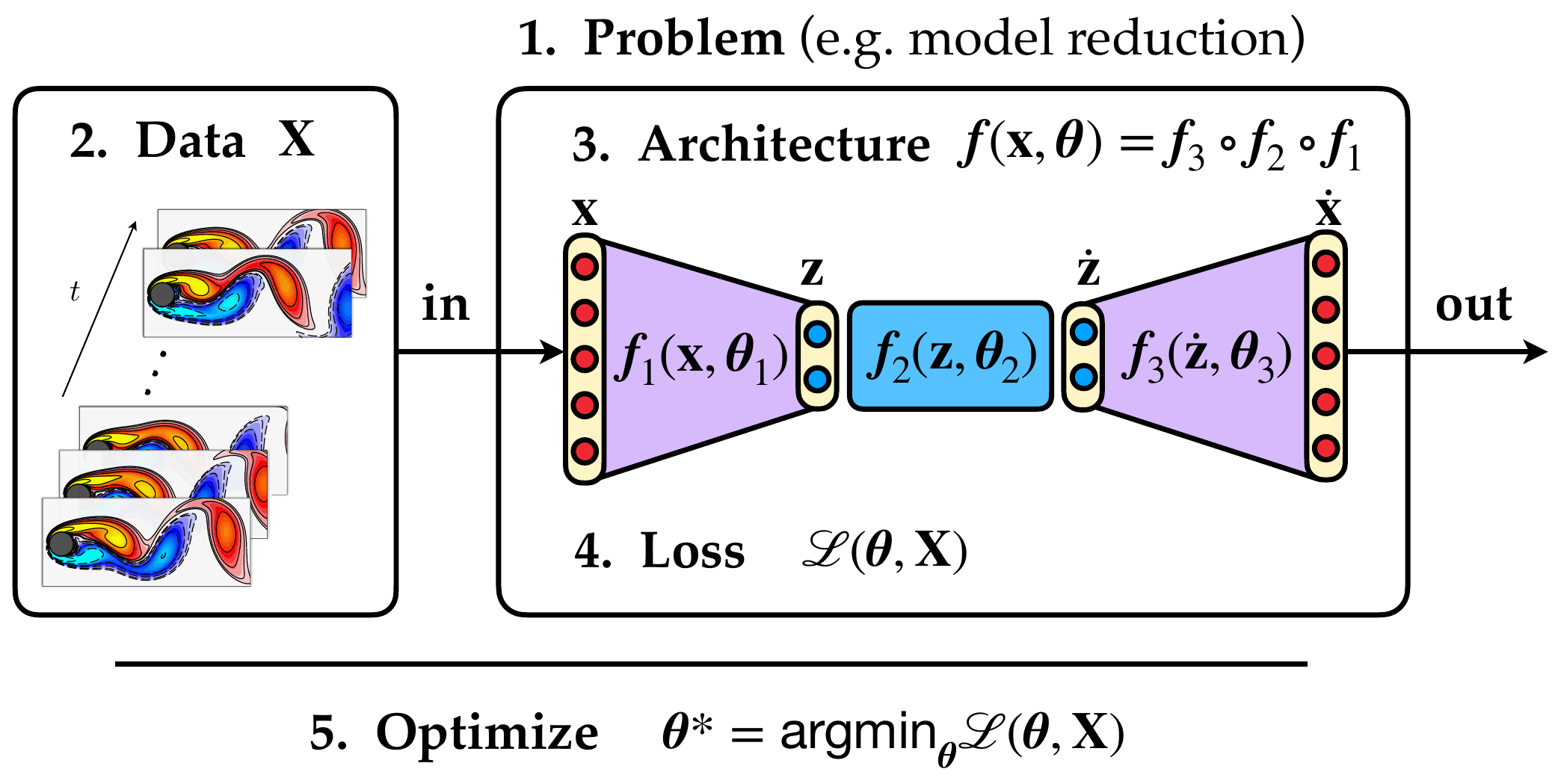}
    \caption{{Schematic of the five stages of machine learning on an example of reduced-order modeling.  In this case, the goal is to learn a low dimensional coordinate system $\mathbf{z} = \boldsymbol{f}_1(\mathbf{x},\boldsymbol{\theta}_1)$ from data in a high-dimensional representation $\mathbf{x}$, along with a dynamical system model $\dot{\mathbf{z}} = \boldsymbol{f}_2(\mathbf{z},\boldsymbol{\theta}_2)$ for how the state $\mathbf{z}$ evolves in time.  Finally, this latent state derivative $\dot{\mathbf{z}}$ must be able to approximate the high dimensional derivative $\dot{\mathbf{x}}$ through the decoder $\dot{\mathbf{x}}\approx\boldsymbol{f}_3(\dot{\mathbf{z}},\boldsymbol{\theta}_3)$.  The loss function $\mathcal{L}(\boldsymbol{\theta},\mathbf{X})$ defines how well the model performs, averaged over the data $\mathbf{X}$.  Finally, the parameters $\boldsymbol{\theta} = \{\boldsymbol{\theta}_1,\boldsymbol{\theta}_2,\boldsymbol{\theta}_3\}$ are found through optimization.}}
    \label{fig01}
\end{figure}
\section{Physics Informed Machine Learning for Fluid Mechanics}

Applied machine learning may be separated into a few canonical steps, each of which provides an opportunity to embed prior physical knowledge: (1) choosing the problem to model or the question to answer; (2) choosing and curating the data used to train the model; (3) deciding on a machine learning architecture to best represent or model this data; (4) designing loss functions to quantify performance and to guide the learning process; and (5) implementing an optimization algorithm to train the model to minimize the loss function over the training data.  
See Fig.~\ref{fig01} for a schematic of this process on the example of reduced-order modeling.
This organization of steps is only approximate, and there are considerable overlaps and tight interconnections between each stage.  
For example, choosing the problem to model and choosing the data to inform this model are two closely related decisions.  
Similarly, designing a custom loss function and implementing an optimization algorithm to minimize this loss function are tightly coupled. 
In most modern machine learning workflows, it is common to iteratively revisit earlier stages based on the outcome at later stages, so that the machine learning researcher is constantly asking new questions and revising the data, the architecture, the loss functions, and the optimization algorithm to improve performance.  
Here, we discuss these canonical stages of machine learning, investigate how to incorporate physics, and review examples in the field of fluid mechanics.  
This discussion is largely meant to be a high-level overview, and many more details can be found in recent reviews~\cite{Duraisamy2019arfm,Brenner2019prf,Brunton2020arfm,brenner2021machine}.

\subsection{The problem} Data science is the art of asking and answering questions with data. 
The sub-field of machine learning is concerned with leveraging historical data to build models that may be deployed to automatically answer these questions, ideally in real-time, given new data.  
It is critical to select the right system to model, motivated by a problem that is both important and tractable.  
Choosing a problem involves deciding on input data that will be readily available in the future, and output data that will represent the desired output, or prediction, of the model.  
The output data should be determinable from the inputs, and the functional relationship between these is precisely what the machine learning model will be trained to capture. 

The nature of the problem, specifically what outputs will be modeled given what inputs, determines the large classes of machine learning algorithms: \emph{supervised}, \emph{unsupervised}, and \emph{reinforcement learning}. 
In supervised learning, the training data will have expert labels that should be predicted or modeled with the machine learning algorithm.  
These output labels may be discrete, such as a categorical label of a `dog' or a `cat' given an input image, in which case the task is one of \emph{classification}. 
If the labels are continuous, such as the average value of lift or drag given a specified airfoil geometry, then the task is one of \emph{regression}.  
In unsupervised learning, there are no expert labels, and structure must be extracted from the input data alone; thus, this is often referred to as \emph{data mining}, and constitutes a particularly challenging field of machine learning. Again, if the structure in the data is assumed to be discrete, then the task is \emph{clustering}. 
After the clusters are identified and characterized, these groupings may be used as proxy labels to then classify new data.  
If the structure in the data is assumed to be continuously varying, then the task is typically thought of as an \emph{embedding} or \emph{dimensionality reduction} task.  
Principal component analysis (PCA) or proper orthogonal decomposition (POD) may be thought of as unsupervised learning tasks that seek a continuous embedding of reduced dimension~\cite{Brunton2019book}. 
Reinforcement learning is a third, large branch of machine learning research, in which an \emph{agent} learns to make control decisions to interact with an environment for some high level objection~\cite{Sutton1998book}.  
Examples include learning how to play games~\cite{mnih2015human,silver2017mastering}, such as chess and go.

\paragraph{Embedding physics:} Deciding on what phenomena to model with machine learning is often inherently related to the underlying physics. 
Although classical machine learning has been largely applied to ``static" tasks, such as image classification and the placement of advertisements, increasingly it is possible to apply these techniques to model physical systems that evolve in time according to some \emph{rules} or \emph{physics}.  
For example, we may formulate a learning problem to find and represent a conserved quantity, such as a Hamiltonian, purely from data~\cite{kaiser2018discovering}.
Alternatively, the machine learning task may be to model time-series data as a differential equation, with the learning algorithm representing the dynamical system~\cite{Schmidt2009science,Schmid2010jfm,Brunton2016pnas,pathak2017using,vlachas2018data}. 
Similarly, the task may involve learning a coordinate transformation where these dynamics become simplified in some \emph{physical} way; i.e., coordinate transformations to linearize or diagonalize/decouple dynamics~\cite{Lusch2018natcomm,Wehmeyer2018jcp,Mardt2018natcomm,Takeishi2017nips,Li2017chaos,Yeung2017arxiv,Otto2019siads,Champion2019pnas}.

\paragraph{Examples in fluid mechanics:} There are many \emph{physical} modeling tasks in fluid mechanics that are benefiting from machine learning~\cite{Brenner2019prf,Brunton2020arfm}.  
A large field of study focuses on formulating turbulence closure modeling as a machine learning problem~\cite{Duraisamy2019arfm,ahmed2021closures}, such as learning models for the Reynolds stresses~\cite{Ling2016jfm,Kutz2017jfm} or sub-gridscale turbulence~\cite{Maulik2019jfm,novati2021automating}. 
{Designing useful input features is also an important way that prior physical knowledge is incorporated into turbulence closure modeling~\cite{Wang2017prf,zhu2019machine,zhu2021turbulence}.} 
Similarly, machine learning has recently been focused on the problem of improving computational fluid dynamics (CFD) solvers~\cite{bar2019learning,Thaler2019jcp,stevens2020enhancement,kochkov2021machine}.  
Other important problems in fluid mechanics that benefit from machine learning include super-resolution~\cite{erichson2020shallow,fukami2019super}, robust modal decompositions~\cite{Taira2017aiaa,Taira2020aiaaj,Scherl2020prf}, network and cluster modeling~\cite{nair2015network,Kaiser2014jfm,fernex2021cluster}, 
control~\cite{maceda2021stabilization,zhou2020artificial} and reinforcement learning~\cite{fan2020reinforcement,verma2018efficient}, and design of experiments in cyberphysical systems~\cite{fan2019robotic}.  
{Aerodynamics is a large related field with significant data-driven advances~\cite{kou2021data}.} 
The very nature of these problems embeds the learning process into a larger physics-based framework, so that the models are more physically relevant by construction.  

\subsection{The data}  Data is the lifeblood of machine learning, and our ability to build effective models relies on what data is available or may be collected.  
As discussed earlier, choosing data to inform a model is closely related to choosing what to model in the first place, and therefore this stage cannot be strictly separated from the choice of a problem above. 
The quality and quantity of data directly affects the resulting machine learning model. 
Many machine learning architectures, such as deep neural networks, are essentially sophisticated interpolation engines, and so having a diversity of training data is essential to these models being useful on unseen data. 
For example, modern deep convolutional neural networks rose to prominence with their unprecedented classification accuracy~\cite{Krizhevsky2012nips} on the ImageNet data base~\cite{deng2009imagenet}, which contains over $14$ million labeled images with over $20,000$ categories, providing a sufficiently large and rich set of examples for training. 
This pairing of a vast labeled data set with a novel deep learning architecture is widely regarded as the beginning of the modern era of deep learning~\cite{Goodfellow2016book}.

\paragraph{Embedding physics:} The training data provides several opportunities to embed prior physical knowledge.  
If a system is known to exhibit a symmetry, such translational or rotational invariance, then it is possible to augment and enrich the training data with shifted or rotated examples.  
More generally, it is often assumed that with an abundance of training data, these physical invariances will automatically be learned by a sufficiently expressive architecture.  
However, this approach tends to require considerable resources, both to collect and curate the data, as well as to train increasingly large models, making it more appropriate for industrial scale, rather than academic scale, research.  
In contrast, it is also possible to use physical intuition to craft new features from the training data, for example by applying a coordinate transformation that may simplify the representation or training. 
{Physical data often comes from multiple sources with different fidelity, such as from numerical simulations, laboratory experiments, and in-flight tests.  This is an important area of research for flight testing and unsteady aerodynamics~\cite{kou2021data}, and recently physics informed neural networks have been used with multifidelity data to approximate PDEs~\cite{meng2020composite}.}

\paragraph{Examples in fluid mechanics:}  Fluids data is notoriously vast and high-dimensional, with individual flow fields often requiring millions (or more!) degrees of freedom to characterize.  
Moreover, these flow fields typically evolve in time, resulting in a time series of multiple snapshots.  
Although vast in the spatial and/or temporal dimensions, data is often rather sparse in parameter space, as it is expensive to numerically or experimentally investigate multiple geometries, Reynolds numbers, etc. 
{Thus there are many algorithms designed for both rich and sparse data.}
Other considerations involve exciting transients and observing how the system evolves when it is away from its natural state.  
In many other cases, fluids data might be quite limited, for example given by time-series data from a few pressure measurements on the surface of an airfoil, or from force recordings on an experimental turbine.  

\subsection{The architecture} Once a problem has been identified, and data is collected and curated, it is necessary to choose an architecture with which to represent the machine learning model.  
Typically, a machine learning model is a function that maps inputs to outputs
\begin{align}
    \mathbf{y} = \mathbf{f}(\mathbf{x};\boldsymbol{\theta})
\end{align}
and this function is generally represented within a specified family of functions parameterized by values in $\boldsymbol{\theta}$.  
For example, a linear regression model would model outputs as a linear function of the inputs, with $\boldsymbol{\theta}$ parameterizing this linear map, or matrix.  
Neural networks have emerged as a particularly powerful and flexible class of models to represent functional relationships between data, and they have been shown to be able to approximate arbitrarily complex functions with sufficient data and depth~\cite{hornik1989multilayer,hornik1991approximation}. 
There is a tremendous variety of potential neural network architectures~\cite{Brunton2019book}, limited only by the imagination of the human designer.  
The most common architecture is a simple feedforward network, in which data enters through an input layer and maps sequentially through a number of computational layers until an output layer. 
Each layer consists of nodes, where data from nodes in the previous layer are combined in a weighted sum and processed through an activation function, which is typically nonlinear.  
In this way, neural networks are fundamentally compositional in nature.  
The parameters $\boldsymbol{\theta}$ determine the network weights for how data is passed from one layer to the next, i.e. the weighted connectivity matrices for how nodes are connected in adjacent layers.  
The overarching network topology (i.e., how many layers, how large, what type of activation functions, etc.) is specified by the architect or determined in a meta-optimization, thus determining the family of functions that may be approximated by that class of network.  
Then, the network weights for the specific architecture are optimized over the data to minimize a given loss function; these stages are described next. 

It is important to note that not all machine learning architectures are neural networks, although they are one of the most powerful and expressive modern architectures, powered by increasingly big data and high performance computing.  
Before the success of deep convolutional networks on the ImageNet dataset, neural networks were not even mentioned in the list of top ten machine learning algorithms~\cite{Wu2008kis}. 
Random forests~\cite{Breiman2001ml} and support vector machines~\cite{Scholkopf2002book} are two other leading architectures for supervised learning. 
{Bayesian methods are also widely used, especially for dynamical systems~\cite{blanchard2021bayesian}.}
Genetic programming has also been widely used to learn human-interpretable, yet flexible representations of data for modeling~\cite{Bongard2007pnas,Schmidt2009science,cranmer2019learning,cranmer2020discovering} and control~\cite{zhou2020artificial}. 
In addition, standard linear regression and generalized linear regression are still widely used for modeling time-series data, especially in fluids.  
The dynamic mode decomposition (DMD)~\cite{Schmid2010jfm,Kutz2016book,Taira2017aiaa} employs linear regression with a low-rank constraint in the optimization to find dominant spatiotemporal coherent structures that evolve linearly in time. 
The sparse identification of nonlinear dynamics (SINDy)~\cite{Brunton2016pnas} algorithm employs generalized linear regression, with either a sparsity promoting loss function~\cite{Tibshirani1996lasso} or a sparse optimization algorithm~\cite{Brunton2016pnas,Zheng2019ieeeacess}, to identify a differential equation model with as few model terms as are necessary to fit the data.

\paragraph{Embedding physics:} Choosing a machine learning architecture with which to model the training data is one of the most intriguing opportunities to embed physical knowledge into the learning process. 
Among the simplest choices are convolutional networks for translationally invariant systems, and recurrent networks, such as long-short-time memory (LSTM) networks~\cite{vlachas2018data} or reservoir computing~\cite{pathak2017using,pathak2018model}, for systems that evolve in time. 
{LSTMs have recently been used to predict aeroelastic responses across a range of Mach numbers~\cite{li2019deep}.}
More generally, equivariant networks seek to encode various symmetries by construction, which should improve accuracy and reduce data requirements for physical systems~\cite{thomas2018tensor,miller2020relevance,wang2020incorporating,batzner2021se}. 
Autoencoder networks enforce the physical notion that there should be low-dimensional structure, even for high-dimensional data, by imposing an information bottleneck, given by a constriction of the number of nodes in one or more layers of the network.  
Such networks uncover nonlinear manifolds where the data is compactly represented, generalizing the linear dimensionality reduction obtained by PCA and POD. 
It is also possible to embed physics more directly into the architecture, for example by incorporating Hamiltonian~\cite{greydanus2019hamiltonian,finzi2020simplifying} or Lagrangian~\cite{cranmer2020lagrangian,zhong2020unsupervised} structure. 
There are numerous successful examples of physics-informed neural networks (PINNs)~\cite{Raissi2019jcp,pang2019fpinns,yang2020physics,mao2020physics,karniadakis2021physics}, which solve supervised learning problems while being constrained to satisfy a governing physical law.  
Graph neural networks have also shown the ability to learn generalizable physics in a range of challenging domains~\cite{battaglia2018relational,cranmer2019learning,sanchez2020learning}.
Deep operator networks~\cite{lu2021learning} are able to learn continuous operators, such as governing partial differential equations, from relatively limited training data.

\paragraph{Examples in fluid mechanics:}  There are numerous examples of custom neural network architectures being used to enforce physical solutions for applications in fluid mechanics.  The work of Ling et al.~\cite{Ling2016jfm} designed a custom neural network layer that enforced Galilean invariance in the Reynolds stress tensors that they were modeling. 
Related Reynolds stress models have been developed using the SINDy sparse modeling approach~\cite{beetham2020formulating,beetham2021sparse,schmelzer2020discovery}. 
{Hybrid models that combine linear system identification and nonlinear neural networks have been used to model complex aeroelastic systems~\cite{kou2019hybrid}.}
The hidden fluid mechanics (HFM) approach is a physics-informed neural network strategy that encodes the Navier-Stokes equations while being flexible to the boundary conditions and geometry of the problem, enabling impressive physically quantifiable flow field estimations from limited data~\cite{raissi2020science}.
{Sparse sensing has also been used to recover pressure distributions around airfoils~\cite{zhao2021research}.} 
The Fourier neural operator is a novel operator network that performs super-resolution upscaling and simulation modeling tasks~\cite{li2020fourier}. 
Equivariant convolutional networks have been designed and applied to enforce symmetries in high-dimensional complex systems from fluid dynamics~\cite{wang2020incorporating}. 
Physical invariances have also been incorporated into neural networks for subgrid-scale scalar flux modeling~\cite{frezat2021physical}.  
Lee and Carlberg~\cite{lee2020model} recently showed how to incorporate deep convolutional autoencoder networks into the broader reduced-order modeling framework~\cite{Noack2003jfm,Benner2015siamreview,Rowley2017arfm}, taking advantage of the superior dimensionality reduction capabilities of deep autoencoders.  

\subsection{The loss function} The loss function is how we quantify how well the model is performing, often on a variety of tasks.  
For example, the $L_2$ error between the model output and the true output, averaged over the input data, is a common term in the loss function. 
In addition, other terms may be added to regularize the optimization (e.g., the $L_1$ or $L_2$ norm of the parameters $\boldsymbol{\theta}$ to promote parsimony and prevent overfitting). 
Thus, the loss function typically balances multiple competing objectives, such as model performance and model complexity. 
The loss function may also incorporate terms used to promote a specific behavior across different sub-networks in a neural network architecture.  
Importantly, the loss function will provide valuable information used to approximate gradients required to optimize the parameters.  

\paragraph{Embedding physics:} Most of the physics-informed architectures described above involve custom loss functions to promote the efficient training of accurate models.  
It is also possible to incorporate physical priors, such as sparsity, by adding $L_1$ or $L_0$ regularizing loss terms on the parameters in $\boldsymbol{\theta}$.  
In fact, parsimony has been a central theme in physical modeling for century, where it is believed that balancing model complexity with descriptive capability is essential in developing models that generalize.  
The sparse identification of nonlinear dynamics algorithm~\cite{Brunton2016pnas} learns dynamical systems models with as few terms from a library of candidate terms as are needed to describe the training data.  
There are several formulations involving different loss terms and optimization algorithms that promote additional physical notions, such as stability~\cite{kaptanoglu2021promoting} and energy conservation~\cite{Loiseau2017jfm}.  
Stability promoting loss functions based on notions of Lyapunov stability have also been incorporated into autoencoders, with impressive results on fluid systems~\cite{erichson2019physics}.

\paragraph{Examples in fluid mechanics:}  Sparse nonlinear modeling has been used extensively in fluid mechanics, adding sparsity-promoting loss terms to learn parsimonious models that prevent overfitting and generalize to new scenarios.  
SINDy has been used to generate reduced-order models for how dominant coherent structures evolve in a flow for a range of configurations~\cite{Loiseau2017jfm,Loiseau2018jfm,loiseau2020data,deng2020low,deng2021galerkin}. 
These models have also been extended to develop compact closure models~\cite{beetham2020formulating,beetham2021sparse,schmelzer2020discovery}.  
Recently, the physical notion of \emph{boundedness} of solutions, which is a fundamental concept in reduced-order models of fluids~\cite{Schlegel2015jfm}, has been incorporated into the SINDy modeling framework as a novel loss function.  
Other physical loss functions may be added, such as adding the divergence of a flow field as a loss term to promote solutions that are incompressible~\cite{wang2020towards}. 

\subsection{The optimization algorithm} Ultimately, machine learning models are trained using optimization algorithms to find the parameters $\boldsymbol{\theta}$ that best fit the training data.  
Typically, these optimization problems are both high-dimensional and non-convex, leading to extremely challenging optimization landscapes with many local minima. 
While there are powerful and generic techniques for convex optimization problems~\cite{Grant2008cvx,Boyd2009book}, there are few generic guarantees for convergence or global optimality in non-convex optimization. 
Modern deep neural networks have particularly high-dimensional parameters $\boldsymbol{\theta}$ and require large training data sets, which necessitate stochastic gradient descent algorithms.  
In a sense, the optimization algorithm is the engine powering machine learning, and as such, it is often abstracted from the decision process. 
However, developing advanced optimization algorithms is the focus of intense research efforts.  
It is also often necessary to explicitly consider the optimization algorithm when designing a new architecture or incorporating a novel loss term.  

\paragraph{Embedding physics:} There are several ways that the optimization algorithm may be customized or modified to incorporate prior physical knowledge.  
One approach is to explicitly add constraints to the optimization, for example that certain coefficients must be non-negative, or that other coefficients must satisfy a specified algebraic relationship with each other. 
Depending on the given machine learning architecture, it may be possible to enforce energy conservation~\cite{Loiseau2017jfm} or stability constraints~\cite{kaptanoglu2021promoting} in this way. 
Another approach involves employing custom optimization algorithms required to minimize the physically motivated loss functions above, which are often non-convex.   In this way, the line between loss function and optimization algorithm are often blurred, as they are typically tightly coupled.  For example, promoting sparsity with the $L_0$ norm is non-convex, and several relaxed optimization formulations have been developed to approximately solve this problem.  
The sparse relaxed regularized regression (SR3) optimization framework~\cite{Zheng2019ieeeacess} has been developed specifically to handle challenging non-convex loss terms that arise in physically motivated problems.

\paragraph{Examples in fluid mechanics:}  Loiseau~\cite{Loiseau2017jfm} showed that it is possible to enforce energy conservation for incompressible fluid flows directly by imposing skew-symmetry constraints on the quadratic terms of a sparse generalized linear regression (i.e. SINDy) model.  
These constraints manifest as equality constraints on the sparse coefficients $\boldsymbol{\theta}$ of the SINDy model.  
Because the standard SINDy optimization procedure is based on a sequentially thresholded least-squares procedure, it is possible to enforce these equality constraints at every stage of the regression, using the Karush–Kuhn–Tucker (KKT) conditions.  
The SR3 optimization package~\cite{Zheng2019ieeeacess} was developed to generalize and extend these constrained optimization problems to more challenging constraints, and to more generic optimization problems.  
This is only one of many examples of custom optimization algorithms being developed to train machine learning models with novel loss functions or architectures.

\section{Parting Thoughts}
This brief paper has attempted to provide a high level overview of the various stages of machine learning, how physics can be incorporated at each stage, and how these techniques are being applied today in fluid mechanics. 
Machine learning for physical systems requires careful consideration in each of these steps, as every stage provides an opportunity to incorporate prior knowledge about the physics. 
A working definition of physics is the part of a model that generalizes, and this is one of the central goals of machine learning models for physical systems. 
It is also important to note that machine learning is fundamentally a collaborative effort, as it is nearly impossible to master every stage of this process.  

The nature of this topic is mercurial, as new innovations are being introduced every day that improve our capabilities and challenge our previous assumptions.  
Much of this work has deliberately oversimplified the process of machine learning and the field of fluid mechanics.  
Machine learning is largely concerned with fitting functions from data, and so it is important to pick the right functions to fit.  
The inputs to the function are the variables and parameters that we have access to or control over, and the outputs are quantities of interest that we would like to accurately and efficiently approximate in the future.  
It is a fruitful exercise to revisit classically important problems where progress was limited by our ability to represent complex functions.  For example, Ling et al.~\cite{Ling2016jfm} had great success revisiting the classical Reynolds stress models of Pope~\cite{pope1975more} with powerful modern techniques. 
More fundamentally, machine learning is about asking and answering questions with data.  
We can't forget why we are asking these questions in the first place:  because we are curious, and there is value in knowing the answer.  

\section*{Disclaimer}
Any omission or oversight was the result of either ignorance, forgetfulness, hastiness, or lack of imagination on my part.  
These notes are not meant to be exhaustive, but rather to provide a few concrete examples from the literature to guide researchers getting started in this field.  
This field is growing at an incredible rate, and these examples provide a tiny glimpse into a much larger effort.  
I have tried to sample from what I consider some of the most relevant and accessible literature. 
However, a disproportionate number of references are to work by my close collaborators, as this is the work I am most familiar with.  
If I have missed any important references or connections, or mis-characterized any works cited here, please let me know and I'll try to incorporate corrections in future versions of these notes.  

\section*{Acknowledgments} 
 SLB  acknowledges many valuable discussions and perspectives gained from collaborators and coauthors Petros Koumoutsakos, J. Nathan Kutz, Jean-Christophe Loiseau, and Bernd Noack. 

 \begin{spacing}{.9}
 \setlength{\bibsep}{6.25pt}

 \end{spacing}
\end{document}